\documentclass[aps,twocolumn,prl,showpacs,color,psfig,epsf]{revtex4}
\usepackage{amsmath}
\usepackage{amsfonts}
\usepackage[dvips]{graphicx}
\begin{document}
                                                                                
\title{Close contact fluctuations: the seeding of signalling domains in the
immunological synapse}
                                                                                
\author{Amit K Chattopadhyay and Nigel J Burroughs}

\affiliation{Mathematics Insitute, University of Warwick, Coventry CV4 7AL, UK}
                                                                                
\begin{abstract}
\noindent
We analyse the size and density of thermally induced regions of close
contact in cell:cell contact interfaces within a harmonic
potential approximation, estimating these regions to be below
${1/10}\:{\mathrm{th}}$ of a micron
across. Our calculations indicate that as the distance between the 
close contact threshold depth and
the mean membrane-membrane separation increases, the density of close contact
patches decreases exponentially while there is only a minimal variation in their mean size.
The technique developed can be used to calculate
the probability of first crossing in reflection symmetry violating systems. \\
\end{abstract}


\maketitle

\noindent 
Surface contact between cells is a key mechanism for information 
transfer in many biological systems. These can be both long 
term or permanent connections as in the neurological synapse, or as 
discovered more recently, transient and highly dynamic as in the 
immunological synapse \cite{syn_rev}.  T-cells (a class of lymphocytes) make 
transient contact with 'target' cells whilst scanning for the presence 
of their specific antigen, antigen recognition resulting in the 
stabilisation of the contact and generation of a macroscopic receptor 
patternation in the contact interface, or a so called immunological synapse \cite{syn_rev}.  A 
fundamental observation is that the contact interface is 
heterogeneous, both in the physical separation of the two cell 
surfaces \cite{revy01} and in the local signalling properties 
\cite{syn_rev,krummel00,freiberg03}.  Differences in the extracellular 
lengths of key molecules/bonds is believed to underpin both these 
processes with a predominant division between short and long bond 
length molecular species.  Of note is that essential antigen 
signalling receptors, such as the T-cell receptor (TCR), are small molecules with a 
ligand-receptor bond length of 14nm (membrane to membrane span)  
\cite{syn_rev}, while an essential phosphatase (CD45), a major 
component of the glycocalyx, has a length of 25-40nm and is not 
believed to have a natural ligand. T cell signalling, or antigen 
detection, thus requires tight cell:cell contact to allow 
TCR binding, whilst such regions necessarily require the 
spatial exclusion of the large molecules comprising the glycocalyx. Spatial 
heterogeneity in the membrane profile within the contact interface 
is therefore essential for the functioning of the cell contact.
Early patterns (50sec) in cell interfaces show random small clusters 
of TCRs \cite{krummel00,freiberg03}, regions where signalling 
intermediaries appear to congregate.  These regions of close contact 
are presumably formed from fluctuations in the initial contact 
surfaces.  At later times signalling appears to
be focused in distinct stable  microclusters \cite{yokosuka05}.
This dependence of signalling on 
spatial heterogeneity introduces 
a key 'exposure' problem; ligand detection requires that 
regions of close contact comprise a significant area within the interface while
they must be sufficiently large that they can be 
stabilised when segregation is energetically favourable \cite{nigel1}. 
We examine the spatial statistics of these regions of close contact 
using a linear stochastic model for thermal fluctuations of the membrane separation.


\begin{figure} 
\includegraphics[width=7.5cm, height=7.5cm,angle=0]{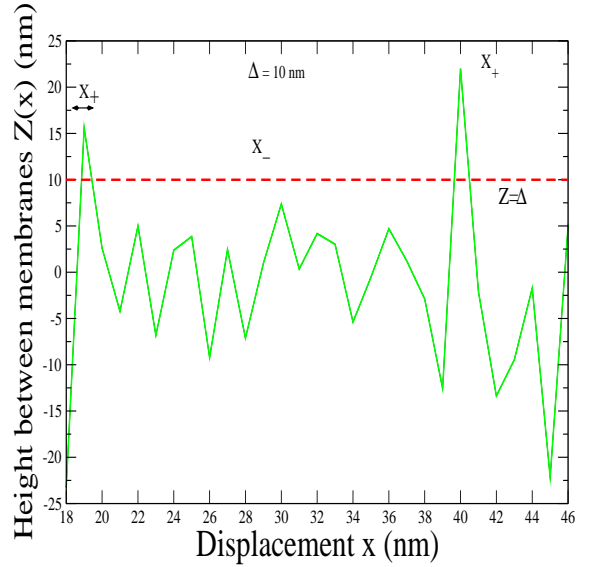}
\caption{A 1D profile of a 2D membrane fluctuating around the line  
$<Z>=0$, from simulated data. $X_{+}$ $\&$ $X_{-}$ are the respective  
sizes of the patches above and below the threshold depth $Z=\Delta=10$nm.} 
\label{fluctuation} 
\end{figure} 
  
\par 
In this letter, our  
 interest is in the size and density of regions of close contact 
(eg membrane-membrane separation $<$20 nm) where effective TCR ligand binding can occur. 
We utilise a  linearised version of the synapse 
reaction-diffusion equations 
\cite{nigel1,chakravarty1}  to model pre pattern dynamics, reducing to a single equation for the
membrane-membrane separation $Z$ around a mean separation (25-50nm) determined by the 
glycocalyx potential and receptor-ligand bond equilibrium. 
In this regime, the fluctuation $Z(\vec x,t)$  
has dynamics  
\begin{equation} 
M \frac{\partial Z}{\partial t} = - B {\nabla}^4 Z + \tau {\nabla}^2 Z  
- \lambda Z + \eta 
\label{equation_of_motion} 
\end{equation} 
where  $B$ is the membrane rigidity, $\tau$ the  
surface tension, $M$ the membrane damping 
constant and $\lambda$ parametrises the rate of relaxation of the membranes close to  
equilibrium, {\it i.e.} the strength of the harmonic approximation to the potential well.  
The thermal noise $\eta(\vec x,t)$ is 
defined using a fluctuation-dissipation relation $<~\eta(\vec x,t) \eta (\vec x',t')>= 
2k_B T M \delta^2(\vec x-\vec x') \delta(t-t')$, $\vec x, \vec x'$ being points in the contact interface.  
The solution $Z(\vec x,t)$ is a Gaussian variate.  
We wish to calculate the probability that the  
displacement
$Z(\vec x,t)$ lies below a 'close contact' threshold  $-\Delta$ 
where $\Delta \sim 5-30$ nm is the membrane-membrane displacement from the mean required
for efficient TCR binding. We identify the region $Z<-\Delta$ as a region of close contact
and determine the 
average size of these close contact patches.
A point to note is the 
symmetry violation of the system around $Z=-\Delta$; specifically the 
average size of a patch above this line (designated by +)  
is not the same as one below this line (designated by $-$). 
The statistics for $Z<-\Delta$ are identical to those for $Z>\Delta$;
thus for presentation we will use $Z> \Delta$ as the threshold.
 
We start by defining the sign (conditional) correlator  for an arbitrary displacement $\vec x$ in the contact interface (relative to the origin) 
\cite{majumdar1,majumdar2}
$A_{+}=< \mathrm{sgn}[Z(\vec x)-\Delta] >_{  Z(\vec 0)> \Delta} $ and $A_{-}=< \mathrm{sgn}[Z(\vec x)-\Delta] >_{  Z(\vec 0)< \Delta}$, $<..>_F$ denoting the average over states where condition $F$ holds.
For simplicity we assume $Z$ is in stationary equilibrium  
and thus initial conditions can be ignored. $Z(\vec 0),\
Z(\vec x)$ define a two variable joint Gaussian  
probability distribution with zero means, variance 
$c_{11}=<Z^2(0)>$ and covariance $c_{12}(\vec x)=<Z(0) Z(\vec x)>$. 
By translational 
symmetry the covariance matrix and $A_\pm$ only depend on the 
spatial displacement $x=|\vec x|$ between the membranes. 
Thus we drop explicit reference to $\vec x$ for simplicity. 
The symmetry relation 
$
A_{+}(x,\Delta)=-A_{-}(x,-\Delta)  
$
means that only $A_{+}$ needs to be evaluated.  
 
An ensemble averaging over the two-variable Gaussian distribution gives 
\begin{eqnarray}     
A_{+}(x)&=& \frac{N_{+}}{\sqrt{2\pi c_{11}}} \int^{\infty}_{-\infty}du\:\: 
sgn(u-\Delta) 
\exp(-\frac{u^2}{2 c_{11}}) \nonumber \\ 
&\times& \int^{\infty}_{(\Delta-u \frac{c_{12}(x)}{c_{11}}){(\frac{c_{11}} 
{{\mathrm{det}}\:c})}^{1/2}} dz \frac{\exp(-z^2/2)}{\sqrt{2\pi}} 
\label{equation1} 
\end{eqnarray} 
where the lower limit follows from the condition $Z(\vec 0)>\Delta$. 
Here $\mathrm{det}\, c=c_{11}^2-c_{12}^2$ and  
the normalisation 
constant $N_{+}$ is defined by the error function  
${N_{+}}^{-1}=\int^{\infty}_{\frac{\Delta}{\sqrt{c_{11}}}} 
du \frac{\exp(-u^2/2)}{\sqrt{2\pi}}$ which is in fact the 
probability of observing a separation 
$Z(\vec x)>\Delta$ at an arbitrary point $\vec x$.  
We define the patch sizes $X_\pm$ for regions where $Z>\Delta, Z<\Delta$ 
respectively (in 2D along an arbitrary vector), Fig. \ref{fluctuation}. 
To evaluate 
$<X_{\pm}>$, we need to evaluate $A_{\pm}'(0,\Delta)$ where the prime 
refers to a derivative with respect to $x=|\vec x|$. 
This follows from the relation $A_{\pm}(x,\Delta)=1-2x/<X_{\pm}>$ as
separation $x \rightarrow 0$, a consequence of the  
fact that the probability of finding a crossing ({\it i.e.} $Z = \Delta$) in a small interval 
of length $x$ is  
$\frac{x}{<X_{\pm}>}$.  
This gives us the exact relation  
$<X_{\pm}>=-\frac{2}{A_{\pm}'(0)}$. 

We proceed to  
compute the derivative as 
$A_{\pm}'(x)={\frac{\partial A_{\pm}}{\partial c_{12}}}. 
{\frac{\partial c_{12}}{\partial x}}$, Eq. (\ref{equation1}) giving 
\begin{equation} 
\frac{\partial A_{+}}{\partial c_{12}} = \frac{{N_{+}}}{\pi}\:\:\frac{\exp[-\frac{{\Delta}^2} 
{2 c_{11}}]}{\sqrt{\mathrm{det}\:\:c}}\:\:\exp[-\frac{c_{11}}{2 \mathrm{det}\:\:c} {\Delta}^2 
{(1-\frac{c_{12}}{c_{11}})}^2] 
\label{Aplus_c12} 
\end{equation} 
 
\noindent 
The 
relevant correlator in 2D is given by 
\begin{eqnarray} 
c_{12}(x) &=& \frac{k_B T}{(2\pi)^2 M} \int d \vec k 
\frac{e^{-i \vec k \cdot \vec x }}{\alpha(\vec k)} \nonumber \\
&=& {k_B T \over 4\pi \sqrt{\lambda B} }
{ K_0 (e^{- \phi /2} \hat x)  -K_0 (e^{ \phi /2} \hat x)  
\over \sinh  \phi}
\label{variance_equations} 
\end{eqnarray} 
 
\noindent 
where $\alpha(\vec k)=\frac{B |\vec k|^4 + \tau | \vec k |^2 + \lambda}{M}$,
$\phi=\log \left(\frac{\tau - \sqrt{\tau^2 - 4 \lambda B}}{2 \sqrt{\lambda B}}   \right)$, $\hat x = \left({\lambda \over B} \right)^{1 \over 4} x$ and $K_0$ is a modified Bessel function of degree 0.  
The final integral uses a Bessel function identity \cite{gradryshik}.
Therefore $c_{11}={k_B T  \phi  \over 4\pi \sqrt{\lambda B} \sinh \phi}$, and 
for small $x$ we find
$A_+(x, \Delta) \sim 1-  \hat x (\log _e \hat x)^{1 \over 2} C$, $C$ a constant.
Thus crossings fail to conform to 
the assumptions above, specifically $c_{12}$ 
is not twice differentiable at $x=0$ \cite{iia}. 
This is a familiar consequence of
Brownian motion crossing behaviour and stems from the
high frequency noise component of $\eta$ that causes repeated crossing
of the threshold in between large excursions away from the threshold.
We regularise the divergence by introducing an infra-red cut-off in
the noise, thus correlator (\ref{variance_equations}) becomes 
\begin{equation}
c_{12}(x) = 
{k_B T \over 2\pi M } \int^{k_m}_0 d k {k J_0(x k) \over \alpha(k)} , \label{c12reg}
\end{equation}
where $J_0$ is a Bessel function of degree 0 and cut-off
$k_m=2 \pi \big/\epsilon$ is 
given by  the smallest length scale $\epsilon$ in the system.
This length scale is on a sub nanometer scale, e.g. the width of lipid molecule head in the membrane.  
A regular expansion for $c_{12}$ at small $x$ now follows

\begin{eqnarray}
&&c_{12} \sim   c_{11} - \nonumber \\
&&  x^2 \left({k_B T  \over 32\pi B} \log_e \left({ Bk_m^4 + \tau k_m^2 +\lambda \over \lambda}  \right) -  {\tau \over 8 B} c_{11}\right)
\label{newequation} 
\end{eqnarray}

\begin{figure} 
\includegraphics[width=7.5cm, height=7.5cm,angle=0]{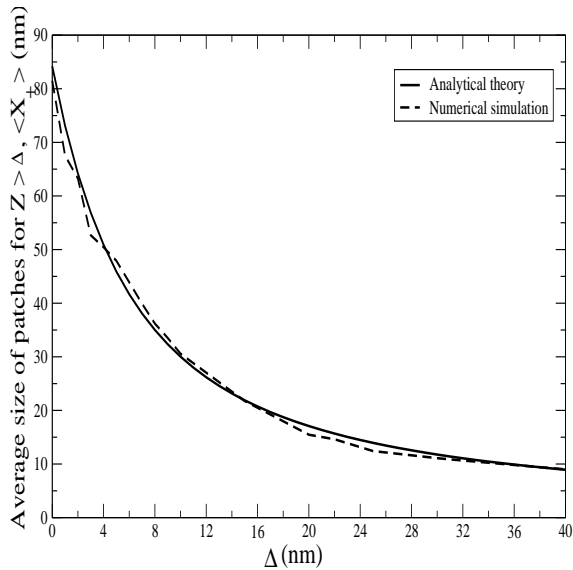}
\caption{Variation of $<X_{+}>$ against $\Delta$ for $Z > \Delta$:  
theoretical estimate from Eq. (\ref{Xequations}) vs numerical simulation 
taken on a lattice, size 1000, spacing 1nm, over 1024 runs. 
A coloured noise
spectrum is used, derived by projection from 2D, giving $<\eta(x,t) \eta (x',t')>= 
2k_B T M s(x-x')\delta^2(\vec x-\vec x') \delta(t-t')$ with 
$ 
s(k)={(\frac{M \alpha(k)}{B})}^{1/4}\:\:\frac{1}{4 \cosh{\frac{1} 
{2}\hat \phi(k)}} ,
$
where $\hat \phi(k) = \log [\frac{\tau + 2 B k^2}{\sqrt{4B M \alpha(k)}} -  
\sqrt{\frac{{(\tau + 2 B k^2)}^2} 
{4 B M \alpha(k)} - 1}]$.} 
\label{width} 
\end{figure}

Provided $\epsilon$ is sufficiently small we have a consistent
regularisation with $c_{12}^{\prime \prime}<0$ at $x=0$. We thus
obtain the following expression
\begin{equation} 
A_{+}'(0,\Delta) = -2 {N_{+} \over f}\:\: {(\frac{\lambda}{B})}^{1/4}\:\:\exp(-{\Delta}^2/{2 c_{11}}) 
\label{Aplus_equation} 
\end{equation} 
where $f=4\pi \sqrt{ \phi \big/ \left( log \left( {k_m^4 B \over \lambda}\right) \sinh  \phi \right)}$ depends only on system parameters. We have retained only the leading order in the cut-off for simplicity.
The mean sizes of the patches above and below the line $Z=\Delta$ now follow, 
\begin{equation} 
<X_{\pm}> ={f \over N_{\pm}} {(\frac{B}{\lambda})}^{1/4} 
\exp({\Delta}^2/{2 c_{11}})  
\label{Xequations} 
\end{equation} 
where the normalisation constant $N_{-}$ is defined as  
${N_{-}}^{-1}=1-{N_{+}}^{-1}$. The dependence on the cut-off is weak
while the length scale is determined by $\sqrt{c_{11} B \big/ k_B T}$.
For a symmetry preserving system with $\Delta=0$  
we have $<X_{\pm}> ={ f\over 2} {(\frac{B}{\lambda})}^{1/4}$. 
Suitable values for the system parameters are \cite{nigel1} 
: $B=11.8\:\:k_B T$, $\tau=5650 k_B T\:\:{\mu m}^{-2}$, 
$M=4.7 \times {10}^6\:\:k_B T\:\:s\:\:{\mu m}^{-4}$ and $\epsilon=1$nm, while 
$\lambda=6.0 \times {10}^5\:\:k_B T\:\:{\mu m}^{-4}$ is approximated 
from the linearised reaction-diffusion equation 
as $\tau \times $ CD45 density, the latter being approximately 100 molecules $\mu m^{-2}$. 
This follows from the force expression in synapse reaction diffusion equations,
$\sum_i \kappa (z-l_i) C_i$, a sum over all molecules $C_i$ that impose a force on the membrane
(bond length $l_i$) with a spring constant $\kappa \sim \tau$, 
\cite{nigel1,kardar}. In early signalling, CD45 will be the dominant component.
These values give $f=2.5$,
$<X_{\pm}(\Delta)>|_{\Delta=0} \sim 84\:\: \mathrm{nm}$. 
The variation of $<X_{\pm}(\Delta)>$ with $\Delta$  
is illustrated in Fig. \ref{width}. Density fluctuations 
in the $C_i$ will causes fluctuations in $\lambda$
which can be included as a
"non-equilibrium temperature"  in Eqn. (1) (fluctuation-dissipation relation);
however this is beyond the current minimalist model.

\par
\begin{figure} 
\includegraphics[height=7.5cm,width=7.5cm,angle=0]{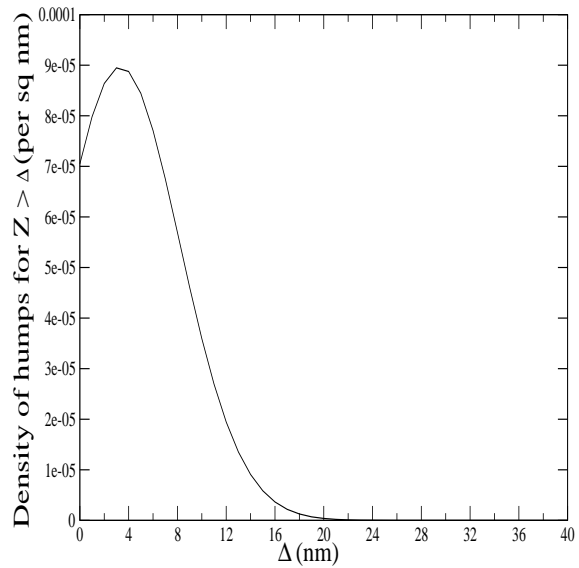}
\caption{Probability density of humps 
around the line $Z=\Delta$ as a function of the average threshold
$\Delta$ (nm) as predicted from Eq. (\ref{hump_equations})}.
\label{probpatch} 
\end{figure}

 
As the threshold $\Delta$ increases above zero the regions 
$Z>\Delta$ develop into isolated patches in 2D. 
We can use the mean size  to estimate the patch density 
${\rho}_{\mathrm{humps}}$ 
by a mean field approximation  
${\rho}_{\mathrm{humps}} {<X_{+}>}^2 = {N_{+}}^{-1}$ 
to obtain  
\begin{equation} 
{{\rho}_{\mathrm{humps}}}^{+} = {N_{+} \over f^2}\:{(\frac{\lambda}{B})}^{1/2}\: 
\exp(-\frac{{\Delta}^2}{c_{11}}) . 
\label{hump_equations} 
\end{equation} 
The 
expected decline in the density of patches as $\Delta$ increases is 
shown in Fig. \ref{probpatch}. For 
large $\Delta \gg c_{11}$ the leading behaviour is $<X_+(\Delta)> \sim 
f {c_{11}^{1 \over 2} \over \sqrt{2 \pi}\Delta} {(\frac{B}{\lambda})}^{1/4} $ 
and $\rho_{\mathrm{humps}}\sim { \Delta \over f^2}
{\left(\frac{2 \pi \lambda}{ c_{11} B}\right)}^{1/2}\exp(-\frac{{\Delta}^2}{2c_{11}})$. These asymptotic approximations capture the contrasting weak 
decline of the width $<X_+>$, and strong decay of the hump density $\rho_{\mathrm{humps}}$ with $\Delta$
in Figs. \ref{width} \& \ref{probpatch}. 
In 1D, regions with $Z>\Delta$ are always disconnected so  
the patch density $\rho_{\mathrm{humps}}$ can be defined 
for all values of $\Delta$.  Further, in 1D, there are no divergences,
whilst in higher dimension the divergences are more severe. These
properties result from the interplay between the 4$^{th}$ order PDE Eq. \ref{equation_of_motion}
and the volume of phase space.

\par 
The probability density function of the distance between crossings 
can also be approximated. This utilises the
probability distribution for the number of
crossings of the line $Z=\Delta$, which is computed by
generalising the traditional 'persistence' analysis \cite{majumdar1,majumdar2}.
We need to discriminate between the two types of crossings, a 
crossing from $Z>\Delta$ to $Z<\Delta$ as $x$ increases, and
the converse.  Let $p^{+}_n(x)$ denote the probability that an
interval of length $x$ contains $n$ crossings of $Z(\vec x,t)$ across
the reference level $\Delta$ when $Z>\Delta$ at the extreme left, and
$p^{-}_n(x)$ is the corresponding probability with $Z<\Delta$ at the
extreme left; in 2D we consider moving a distance $x$ along a
specified vector. 
Then under an independent interval approximation, for $n
\ge 1$, their Laplace transforms have the forms
\begin{eqnarray} 
{\tilde p}^{\pm}_{n}(s)&=& \frac{N_\pm}{2}\frac{(1-{\tilde P}_{+}) 
(1-{\tilde P}_{-}) \left({\tilde P}_{+}\:\:{\tilde P}_{-}\right)^{\frac{n-1}{2}}}{s^2 <X>}, 
\nonumber \\ 
&=&\frac{N_\pm}{2}\frac{{(1-{\tilde P}_{\pm})}^2\:\: 
{\tilde P}_\mp  \left({\tilde P}_{+} {\tilde P}_{-}\right)^{\frac{n-2}{2}}}{s^2 <X> },
\end{eqnarray} 
 
\noindent 
for $n$ odd and even respectively.
Here $P_{\pm}(x)$ is the probability density for the 
distance $x$ between crossings ($+:Z> \Delta$, $-:Z< \Delta$),
and $<X>=(<X_+>+<X_->)\big/2$ is the average distance between consecutive crossings
(any type).
Using the identities  
$\sum^{\infty}_{n=0}\:\:p^{\pm}_{n}(X)=1$, we can now show that 
${\tilde p}^{\pm}_{0}=s^{-1}-N_\pm (1-\tilde P_\pm)\big/(2 s^2 <X>)$, which
agrees with \cite{majumdar2} when $\Delta=0$.
Employing the identity 
$A_{\pm}(x)=\sum^{\infty}_{n=0}\:\:{(-1)}^{n}\:\:p_{\pm}(x)$, we  arrive at two  
coupled  equations relating $P_\pm$ and $A_\pm$,
 
\begin{eqnarray} 
{\tilde A_{\pm}}(s)&=&\frac{1}{s} - \frac{N_\pm}{s^2 <X>} \ 
\frac{
(1-{\tilde P_{-}})
(1-{\tilde P_{+}})}
{1-{\tilde P_{+}}{\tilde P_{-}}}  \ \ 
\label{probability_equations} 
\end{eqnarray} 

\noindent 
Solving these equations then gives the desired pdfs $P_\pm(x)$.

\par 
To summmarize, for the harmonic potential membrane model we have an
exact analytic calculation for the mean size of close contact patches,
$<X_+(\Delta)>$, our calculations suggesting that these are on the
scale of tens of nm. The scale is primarily determined by the
combination $\sqrt{ c_{11} B \big/k_B T}$ and has a leading order behaviour going as $1
\big/ \Delta$ for large $\Delta $. This small patch size
implies that multiple receptor bindings are unlikely within a patch
and close contact patches are unobservable by traditional light
microscopy.  The small size also implies that phosphatase exclusion
(CD45) probably results from density fluctuations, ie a specific
exclusion mechanism is not required in contrast to that needed at
larger sizes \cite{nigel1,chakravarty1}.  The density of patches
decays rapidly with the threshold $\Delta$ on a length scale of
$\sqrt{c_{11}} \sim 5.4nm$, Eq. (8), and indicates that cell membranes must be
highly flexible otherwise the glycocalyx would impose too large a
barrier to allow formation of close contact regions ($\lambda$
increasing with membrane elasticity).  In particular, the glycocalyx
cannot be too deep relative to the size of the TCR ligand-receptor
bond length (14nm) since otherwise the density of patches becomes too
small for antigen detection. The probability of 
T cell signalling depends on the ability of  the TCR to bind it's ligand
and is thus crucially dependent on 
the area of
close contact regions within the cell:cell interface which varies as
$N_+^{-1}$, and on the size of those close contact  patches. There is
an enhancement in triggering as patch sizes increase above 150nm 
\cite{burr06}; thus our estimates suggest that early signalling
relies on patches below this size and enhancement effects only occur
upon aggregation and stabilisation of clusters as the immunological
synapse forms. Such conclusions are somewhat reminiscent of 
\cite{kardar} where formation of a synapse was related to a 
critical value of the system parameters (albeit without 
evaluating the patch size).
We, however, go beyond such qualitative predictions. Our 
calculations clearly suggest that
the membrane correlation length is a determining factor in the area of 
close contact regions in the interface,
which with our parameters limits the threshold to $\Delta <16$ nm.

We thank S.N. Majumdar for helpful discussions. AKC is funded by BBSRC 
grant 88/E17188.

\end{document}